\documentstyle[11pt,aaspp4]{article}

\begin{document}

\title{
	The fading of young stellar populations and the\\
	luminosity functions of dwarf, irregular and starburst galaxies
}
\author{
	David W. Hogg \&
	E. S. Phinney
}
\affil{\sl
Theoretical Astrophysics, California Institute of Technology,\\
mail code 130-33, Pasadena CA 91125 \\
{\tt hogg, esp@tapir.caltech.edu}
}

\begin{abstract}
Dwarf, irregular and infrared-luminous starburst galaxies are all
known to have ``steep'' luminosity functions with faint-end behavior
roughly $\phi(L)\propto L^{-1.8}$.  This form is exactly what is
expected if the luminosities of these objects fade with time as
$L\propto t^{-1.3}$, because the objects spend more time at low
luminosities than high, even if they form with a wide range of initial
masses.  Models of young stellar populations show this fading behavior
when the star formation has occured in a single, short, recent burst.
Steep luminosity functions therefore do not require steep mass
functions if the galaxies are powered by fading bursts.  The
local-galaxy H$\alpha$ luminosity function---which is less steep than
$L^{-1.8}$---is also well-fit by this mechanism, because ionizing
photon flux fades much more quickly than broad-band optical
luminosity.  An age-luminosity relation and a wavelength-dependence of
the luminosity function are both predicted.  In the context of this
mechanism, the slope of the luminosity function provides a constraint
on the stellar initial mass function in the bursts.
\end{abstract}


\section{Introduction}

While the normal field galaxy luminosity function (GLF) $\phi(L)$
(number density per unit luminosity) or $\phi(\log L)$ (number density
per unit log luminosity) is ``flat'' in optical bandpasses at the
faint end, i.e., $\phi(L)\propto L^{-1}$ or $\phi(\log L)={\rm
constant}$ (Efstathiou, Ellis \& Peterson 1988; Loveday et al 1992;
Mobasher, Sharples \& Ellis, 1993; Marzke et al 1994a; Lin et al 1996,
1997; Gardner et al 1997; Ratcliffe et al 1997), many studies have
found that objects in which the luminosity is thought to be dominated
by young stars show a ``steep'' GLF, with roughly $\phi(L)\propto
L^{-1.8}$ at the faint end.  Parameterizing $\phi(L)\propto
L^{\alpha}$, the $60~\mu{\rm m}$ GLF from the IRAS Bright Galaxy
Sample appears to show $\alpha=-1.8$ at the faint end (Soifer et al
1987; although see Saunders et al 1990) despite the fact that these
same objects lie in the flat part of the optical GLF.  The $60~\mu{\rm
m}$ luminosity is thought to originate in dust heated by the radiation
from young stars at ultraviolet wavelengths where dusty galaxies are
optically thick.  Although the faint end of the optical GLF is flat,
there may also be a small ``upturn'' at the very faintest end, around
absolute magnitude $M_B=-16$~mag (Marzke et al 1994a; Driver \&
Phillips 1996; Loveday 1997), which is explained by a luminosity
function with $\alpha=-1.8$, among dwarf (i.e., low-luminosity)
galaxies.  In the case of the CfA survey, the upturn can be explained
entirely by the luminosity function of the Sm-Im galaxies (identified
on the basis of morphology) which show $\alpha=-1.87\pm0.2$ (Marzke et
al 1994b).  Local galaxies spectrally classified as strongly or
recently star-forming also show a steep luminosity function (Heyl et
al 1997).  A steep upturn at the faint end of the GLF is observed for
dwarf galaxies in rich clusters with $\alpha$ ranging from $-1.4$ to
$-2.2$ (Sandage, Binggeli \& Tammann 1985; Driver et al 1994b;
Bernstein et al 1995; De~Propris et al 1995; Lobo et al 1996; Wilson
et al 1997).  A recent measurement of the luminosity function of dwarf
galaxies or ``knots'' formed in the tidal tails of merging galaxies
finds $\alpha=-1.75\pm0.27$ in the $R$-band for these objects
(Hunsberger, Charlton \& Zaritsky 1996).  Compact ``super star
clusters'' observed in the vicinity of starburst galaxies or galaxy
mergers and interpreted as the progenitors of globular clusters show a
luminosity function consistent with $\alpha=-1.8$ although the numbers
are small (Lutz 1991; Holtzman et al 1992; Whitmore et al 1993; Conti
\& Vacca 1994).  Finally, a luminosity function with faint end
behavior $\alpha\approx -1.8$ is often invoked as a natural
explanation of faint galaxy counts and redshift distributions
(Broadhurst, Ellis \& Shanks, 1988; Eales 1993; Koo, Gronwall \&
Bruzual 1993; Driver et al 1994a; Treyer \& Silk 1994; Metcalfe et al
1995; Smail et al 1995; Lilly et al 1995; Ellis et al 1996).  In these
studies, the steep GLF is largely required to account for the large
numbers of faint blue galaxies, which are mainly irregulars
(Glazebrook et al 1995a; Driver et al 1995; Abraham et al 1996) and
thought to have luminosities dominated by young stars.

The steep faint end of the GLF is usually attributed to a steep
underlying galaxy mass function.  A steep mass function at small halo
mass $M_h$ is natural for cold and mixed dark matter models. In the
Press \& Schechter (1974) formalism, on small mass-scales
$\phi(M_h)\,dM_h\propto M_h^{-((9-n)/6)}\,dM_h$, where the
post-recombination power spectrum of density fluctuations has
$P(k)\propto k^n$, with $n\rightarrow -3$ for adiabatic fluctuations
on small scales.  Thus $\phi(M_h)\,dM_h\propto M_h^{-2}dM_h$.  This
has been amply verified by numerical simulation for both cold
(Brainerd \&\ Villumsen 1992) and mixed (Ma \&\ Bertschinger 1994)
dark matter halos.  However, the ejection of gas by early generations
of stars in shallow potential wells implies that the mass converted to
stars rises faster than linearly with $M_h$, so for identical stellar
populations, the galaxy luminosity function should be shallower than
the halo mass function (see Silk \& Wyse 1993 for a review).
Furthermore, in the IR-luminous galaxy sample, the large scatter in
optical-IR colors (Soifer et al 1987) and the lack of correlation
between IR luminosity and galaxy mass inferred from rotation curves
(Lehnert \& Heckman 1996) suggest that the starburst GLF is not
strongly tied to the host galaxy mass function.

In this {\sl Letter,} we remark that there is a natural mechanism
which ensures a steep GLF among young objects: Even if the galaxy mass
function (where now by ``galaxy mass'' is meant ``the mass of that
part of galaxy's baryonic mass which is turned into stars'') is flat,
a GLF of roughly the form $\phi(L)\propto L^{-1.8}$ will be observed
among any population of objects whose luminosities are dominated by
light from short, recent bursts of star formation with a Salpeter-like
initial mass function.  This is because their luminosities decrease
with time in such a way that they spend more time (and are therefore
more numerous) at low luminosities than at high luminosities.  This
kind of mechanism underlies the theoretical explanation of the
$60~{\rm \mu m}$ GLF by Scoville \& Soifer (1991) and a discussion of
the super star cluster luminosity function by Meurer (1995).  An
important feature of this mechanism is that steep mass functions are
not required to explain steep luminosity functions.

\section{Exposition}

The luminosity $L$ of a main sequence star of mass $M>1M_\odot$ scales as
$L\propto M^\eta$, where $\eta=3.9$ (Kippenhahn \&\ Weigert 1990).
So the main sequence lifetime $\tau\propto M/L\propto M^{1-\eta}$.
Let an ensemble of stars be formed in a single burst with initial
mass function $\phi(M)\,dM\propto M^{-(1-x)}\,dM$ (where $x=1.35$ is
the Salpeter slope). The stages of stellar evolution beyond
the main sequence have lifetimes and an integrated consumption of
nuclear fuel less than or comparable to the main sequence values
(Renzini \&\ Buzzoni 1986). Thus for $x<\eta$, the total luminosity
of the evolving ensemble is dominated by stars just leaving
the main sequence, with mass $M=M_{\rm turn}(t)$:
\begin{equation}
L_{\rm TOT}\simeq C\int_{1\,M_{\odot}}^{M_{\rm turn}(t)}\phi(M)L(M)\,dM
\simeq \left. \frac{C}{\eta-x} M\phi(M)L(M)\right\vert_{M_{\rm turn}(t)}\;,
\label{Ltotdef}
\end{equation}
where $C$ is a factor of order 2 (Renzini \&\ Buzzoni 1986)
which accounts for the post-main sequence fuel consumption.
Since $M_{\rm turn}(t)$ is defined by $t=\tau(M_{\rm turn})$,
we have for $x<\eta$,
\begin{equation}
L_{\rm TOT}\propto t^{\zeta},\quad\zeta=\frac{x-\eta}{\eta-1},
\end{equation}
so in this approximation $\zeta\approx-0.9$ for a Salpeter IMF.  The
luminosity in a particular wavelength band $X$ will have a somewhat
different dependence on age than the bolometric luminosity, depending
on bolometric correction, but since stars consume about as much fuel
in the (red) post-main seqence as during the (blue) main sequence, the
dependence in optical and near IR bands will not be vastly different
from that of the bolometric luminosity.

If in wavelength band $X$ an object has luminosity $L_X$ which varies
with time $t$ since its birth as
\begin{equation}
L_X\propto t^{\zeta_X}
\label{zetaXdef}
\end{equation}
and the object is observed at a random times after its birth, the
probability that it is measured to have luminosity $L_X$ is
proportional to $(dL_X/dt)^{-1}\propto L_X^{(1/\zeta_X)-1}$.
Therefore, a population of identical objects of this type born at a
fairly uniform rate over a time interval will appear to have a
luminosity function with faint end slope $\alpha_X=(1/\zeta_X)-1$
despite being intrinsically identical.  The luminosity function is
produced by a spread in ages rather than a distribution of intrinsic
properties.  To make a luminosity function with $\alpha\approx -1.8$,
$\zeta\approx -1.3$ is required.

In any realistic scenario, the objects will have a distribution of
intrinsic sizes or starburst masses.  Consider objects with
luminosities $L_X= M\,\Lambda_X(t)$ where $M$ is the mass of stars
created in the starburst, the objects are selected from a distribution
$\phi(M)$ of starburst masses, and $\Lambda_X(t)\propto t^{\zeta_X}$
is the light per unit mass in wavelength band $X$ as a function of
time.  The faint end of the luminosity function will still show
$\phi(L)\propto L^{(1/\zeta)-1}$ as long as $\phi(M)$ is flatter than
$\phi(M)\propto M^{(1/\zeta)-1}$ (i.e., the exponent is less negative
than $(1/\zeta)-1$) at the small-mass end and steeper than
$\phi(M)\propto M^{(1/\zeta)-1}$ at the large-mass end.  Both
conditions hold if the ``flat'' Schechter function observed for the
local GLF represents the intrinsic mass distribution for galaxies.

With models much more detailed than the scaling argument given at the
beginning of this section, Leitherer \& Heckman (1995) and Bruzual \&
Charlot (1993) predict the fading with time of young stellar
populations in a number of photographic bands and for several models
of the stellar initial mass function (IMF) and metallicity, to which
power laws have been fit.  The results are given in
Table~\ref{tab:zetas}.  The exponents in the ultraviolet and optical,
for Salpeter IMF (slope $x=1.35$) are in the range $-1.4<\zeta<-0.9$,
which lead to luminosity functions with $-2.1<\alpha<-1.7$.  A
Miller-Scalo IMF (slope $x=2.3$) produces slower fading and therefore
steeper (more negative $\alpha$) luminosity functions.  Thus if the
luminosity function of faint galaxies is indeed produced by fading of
young populations, the slope of the luminosity function can be used to
constrain the stellar IMF.  The different luminosity function slopes
observed in different clusters (Binggeli, Sandage \& Tammann 1985;
Driver et al 1994b; Bernstein et al 1995; De~Propris et al 1995; Lobo
et al 1996; Wilson et al 1997) could reflect an environmental
dependence in the stellar IMF, although the statistics are small
enough that the various studies appear to be at least marginally
consistent with one another.

In addition, the evolution of the rate $\Gamma_i$ of production of
ionizing photons (i.e., photons shortward of $912$~\AA) is predicted
(Leitherer \& Heckman, 1995), with roughly $\Gamma_i\propto t^{-4.9}$.
If the strength of H$\alpha$ emission is proportional to $\Gamma_i$,
this fading behavior naturally produces an H$\alpha$ luminosity
function with faint-end slope $\alpha=-1.2$, consistent with a recent
measurement of $\alpha=-1.3\pm 0.2$ for the local H$\alpha$ GLF
(Gallego et al 1995).

As an illustration, Fig.~1 shows: {\sl (a)}~fading laws of the form
\begin{eqnarray}
\Lambda_{\rm bol}(t) = & 0 & \;\;\mbox{for}~t<t_0 \nonumber \\
 & \Lambda_0\,\left(\frac{t}{t_0}\right)^{\zeta} &  \;\;\mbox{for}~t>t_0
\label{powerlaw}
\end{eqnarray}
with $\zeta=-1.4$ and $-0.7$ which serve as approximations to Fig.~7
of Leitherer \& Heckman (1995); {\sl (b)}~distributions of starburst
masses $\phi(\log M)$ (number per decade in mass) of the Schechter
(1976) form
\begin{equation}
\phi(\log M)\propto M\,\phi(M)\propto
 \left(\frac{M}{M^{\ast}}\right)^{\beta+1}\,e^{-M/M^{\ast}}
\end{equation}
where $\beta$ sets the slope of the mass function at the low-mass end
and $M^{\ast}$ (fixed at $10^3~M_{\odot}$ for definiteness) sets the
high-mass cutoff; {\sl (c)}~the resulting luminosity functions
$\phi(\log L)\propto L\,\phi(L)$ (number per decade in luminosity)
under the assumption that the populations are observed at random ages
as they fade; and {\sl (d)}~the age distributions $\phi(\log t)$
(number per decade in age) at two different luminosity levels
($10^{11.5}$ and $10^{10.5}~L_{\odot}$ in this case) .  This shows
that the luminosity function has the predicted power-law form even if
there is a broad distribution of starburst masses.  It also shows that
the distribution of ages can be broad but does indeed get older as the
objects get fainter.  The age function is broadest in the cases in
which the mass function is almost as steep as the luminosity function.

As an illustration, the upturn in the local GLF at absolute magnitude
$M_B\approx -16$ ($H_0=100~{\rm km\,s^{-1}\,Mpc^{-1}}$) found by
Marzke et al (1994a) can be analyzed in terms of the fading mechanism.
The GLF has amplitude $\phi(\log L)\approx 0.09~{\rm Mpc^{-3}}$ at
absolute magnitude $M_B=-16$ or luminosity $L_B=3.7\times
10^8\,L_{\odot}$ and slope $\alpha\approx -2$ faintward.  To order of
magnitude, by dimensional analysis,
\begin{equation}
\phi(\log L_B)\equiv (\ln 10)\,L_B\,\frac{dN}{dV\,dL_B}
 = (\ln 10)\,L_B\,\frac{dN}{dV\,dt}\,\frac{1}{M_{\rm typ}}
 \,\left|\frac{d\Lambda_B}{dt}\right|^{-1}
\end{equation}
where $dN/(dV\,dt)$ is the birth rate density (number per unit time
per unit volume), $M_{\rm typ}$ is the typical starburst mass, and
$d\Lambda_B/dt$ is the rate of change of the light-to-mass ratio in
the $B$ band, evaluated at the light-to-mass ratio
$\Lambda_B=L_B/M_{\rm typ}$ at which starbursts of mass $M_{\rm typ}$
have luminosity $L_B$.  At this value,
\begin{equation}
\frac{d\Lambda_B}{dt}
 =\frac{\zeta_B\,\Lambda_0}{t_0}
  \,\left(\frac{t}{t_0}\right)^{\zeta_B-1}
 =\frac{\zeta_B\,\Lambda_0}{t_0}
  \,\left(\frac{L_B}{M_{\rm typ}\,\Lambda_0}\right)^{1-(1/\zeta_B)}
\end{equation}
where $\Lambda_0=5.8\times 10^8\,L_{\odot}/M_{\odot}$, $t_0=3\times
10^6~{\rm yr}$ and $\zeta_B=-1.0$ are the appropriate values for the
fading-law parameters in equation~(\ref{powerlaw}) for the $B$ band
(Leitherer \& Heckman 1995).  Plugging in, a relationship between the
birth rate density and the typical mass is derived
\begin{equation}
\frac{dN}{dV\,dt}= \frac{|\zeta_B|\,\phi(\log L_B)}{(\ln 10)\,t_0}
   \,\left(\frac{M_{\rm typ}\,\Lambda_0}{L_B}\right)^{1/\zeta_B}
 = 8.3\times 10^{-3}
    \,\left(\frac{M_{\rm typ}}{10^3\,M_{\odot}}\right)^{-1}
    ~{\rm Mpc^{-3}\,Gyr^{-1}}
\end{equation}
for $H_0=100~{\rm km\,s^{-1}\,Mpc^{-1}}$.  Note that the model is not
unique because neither the birth rate density nor the typical mass can
be determined from the above information alone, only a relationship
between the two.

Of course this fading mechanism requires a fairly uniform birth rate.
In the extreme, if in some population all starbursts occured
simultaneously at some moment in the past, no spread of ages would be
observed, and the luminosities would be proportional only to the
intrinsic sizes of the starbursts.  The GLF would have nothing to do
with the fading exponent $\zeta_X$ at all.  If the birth rate is
varying slowly with time, it will affect the final luminosity function
shape.  For instance, if the birth rate is decreasing with time, there
will be more old galaxies than young and therefore more low-luminosity
systems than would be predicted for constant birth rate and a steeper
GLF will be observed (i.e., $\alpha$ will be more negative than
$(1/\zeta_X)-1$).  Again, different slopes among, say, different
galaxy cluster GLFs could indicate different dependences of starburst
birth rate on time.  However, over the small dynamic ranges (and hence
age ranges) over which such GLFs have usually been measured, the
assumption of relatively constant birth rate is not unreasonable.

If individual galaxies undergo multiple starbursts, this fading
mechanism will still ensure a steep GLF, as long as the sources are
fading according to $\Lambda_X(t)\propto t^{\zeta_X}$.  For a
multiple-burst source this will be true at young ages when the
recently formed stars still dominate the light.  However, at late
times the underlying population of old stars from previous bursts will
eventually become a significant, non-fading contribution to the
luminosity.  So multiple-burst sources will show the steep GLF, but
perhaps over a somewhat restricted range of age and therefore
luminosity.  Of course because young stellar populations are so much
brighter per unit mass than old, this will not be a big restriction
unless the sources burst very frequently (in which case the basic
assumption of the model breaks down).

\section{Prediction}

This fading model for the faint end of the luminosity function makes
three important predictions.  The first prediction is that the bright
members of the populations which show these steep luminosity functions
should be mostly young.  The distribution of ages and in particular
the modal age $t_m(L_X)$ at each luminosity $L_X$ in band $X$ should
obey the law $t_m(L_X)\propto L^{1/\zeta_X}$.  A correlation between
dust temperature and luminosity is observed in IRAS galaxies (Miley,
Neugebauer \& Soifer 1985; Rieke \& Lebofsky 1986; Soifer et al 1987).
In models in which the far-infrared emission is from dust heated by
young stars, this relation can be interpreted as an age-luminosity
relation, although it is also naturally produced if the dust is heated
by other mechanisms (e.g., nuclear activity or continuous star
formation).  Spectra of IRAS galaxies in the $\phi(L)\propto L^{-1.8}$
part of the $60~\mu{\rm m}$ luminosity function could be used to
confirm this interpretation of the temperature-luminosity relation.
In the optical an age-luminosity relation for a starburst can be
converted into a color-luminosity relation; the expectation is that
fainter galaxies will be redder, in possible contradiction to the
evidence from some samples of extragalactic HII regions (Telles 1995)
and faint galaxy counts (e.g., Koo \& Kron 1992).  Again, spectra of
nearby dwarf, irregular or infrared-luminous galaxies could in
principle be used to determine a quantitative age-luminosity relation.
Because the luminosities are determined more by age than by intrinsic
properties, there ought to be little correlation between the
luminosities and dynamical masses of the galaxies, as measured from
rotation curves or velocity dispersions.  Of course the width of the
age distribution at a given luminosity depends on the underlying
starburst mass distribution discussed above.  If the starburst mass
distribution $\phi(m)$ is a Schechter function with faint end slope
$\beta$ the width of the age distribution increases as
$\beta\rightarrow\alpha=(1/\zeta)-1$; see Fig.~1{\sl (d)}.  (Recall
that if $\beta<\alpha$, the overall luminosity function will have
faint-end slope $\beta$ rather than $\alpha$.)  In principle this
effect could be used to constrain the distribution of starburst
masses.

The luminosity function shape depends on the objects fading out of the
sample with time, so the second prediction is that this mechanism can
only explain the slope of the luminosity function over as many
magnitudes as the fading obeys the power law.  This range appears to
be at least $7.5$~mag (factor of $10^3$) for the visible-band
luminosities (Bruzual \& Charlot, 1993)---unless the decay is
interrupted by a subsequent burst of star formation, of course---but
luminosity functions with steep faint-end slopes over a significantly
greater range cannot be explained entirely by this mechanism.  The
largest dynamic range for which the slope has been measured so far is
in the IRAS sample which appears to show slope $\alpha=-1.8$ over a
factor of roughly $300$ in $60~\mu{\rm m}$ luminosity (Soifer et al
1987).  The luminosity functions of cluster galaxies will have to be
measured to levels more than three magnitudes fainter than current
observational limits in order to exceed the allowed range.

The third prediction comes from the different slopes of $\Lambda(t)$
in different wavelength bands.  These translate into different
predicted luminosity function slopes $\alpha$.  In general, the
luminosity function is expected to become steeper ($\alpha$ more
negative) as the objects are observed in longer-wavelength bands.
This prediction is related to the color-luminosity relation predicted
above and is easy to test with multi-band imaging of dwarf galaxy
populations.  There is already some evidence in contradiction to this
prediction for galaxy cluster GLFs (Wilson et al 1997), and faint
galaxy counts, which may be naturally explained with steep luminosity
functions, become flatter, not steeper, with increasing wavelength
(e.g., Koo \& Kron 1992).  Of course the faint galaxy counts are
difficult to interpret, since the observed objects are effectively
integrated over a range of redshifts (and hence rest wavelengths) and
there is the possibility that multiple distinct populations play
comparable roles.  It is encouraging that the H$\alpha$ luminosity
function (Gallego et al 1995) is well-fit by this mechanism; under
this interpretation the H$\alpha$ luminosity becomes more a measure of
time since most recent starburst than a measure of star-formation
rate, as is conventionally assumed.  Conclusions about the volume
averaged star-formation rate of the local Universe (Gallego et al
1995) are not strongly affected by this change in interpretation,
however, because such conclusions only depend on total emission of
ionizing photons, not whether or not the emission is simultaneous with
or subsequent to the star formation.

\section{Application}

Finally, we emphasize that if the fading of short, recent bursts of
star formation does indeed explain the faint end slope of galaxy
luminosity functions, the measured slope becomes a strong constraint
on the stellar IMF in these objects.  Since other techniques (e.g.
population synthesis models) often do not constrain the IMF uniquely
in individual objects (e.g., Santos et al 1995), the slope of the
faint end of the galaxy luminosity function may become an extremely
useful diagnostic for star and galaxy formation models.

\acknowledgements
We benefitted from helpful comments from Lee Armus, Roger Blandford,
Tim Heckman, Jeremy Heyl, Nick Scoville, Tom Soifer and an anonymous
referee.  Claus Leitherer, Tim Heckman and Jeff Goldader generously
provided us with published data in electronic form.  Support from the
National Science Foundation is gratefully acknowledged (DWH under
AST-9529170 and ESP under AST-9315455).


\newpage
\begin{deluxetable}{lccccc}
\footnotesize
\tablecaption{Exponents of Luminosity Decay\tablenotemark{a} \label{tab:zetas}}
\tablewidth{0pt}
\tablehead{
\colhead{\null} & \colhead{$\zeta_{\rm bol}$} & \colhead{$\zeta_U$} &
\colhead{$\zeta_B$} & \colhead{$\zeta_V$} & \colhead{ $\zeta_K$}
}
\startdata
Ref 1\tablenotemark{b} : $x=1.35$, $Z=Z_\odot$    & ---    & $-1.2$ & $-1.0$ & $-0.9$ & $-0.7$\nl
Ref 2\tablenotemark{c} : $x=1.35$, $Z=0.1Z_\odot$ & $-1.4$ & $-1.2$ & $-1.0$ & $-1.0$ & $-0.5$\nl
Ref 2\tablenotemark{c} : $x=2.3$, $Z=0.1Z_\odot$  & $-0.7$ & $-0.5$ & $-0.4$ & $-0.3$ & $-0.0$\nl
\enddata
\tablenotetext{a}{See equation (\protect\ref{zetaXdef}).}
\tablenotetext{b}{Ref 1: Bruzual \&\ Charlot 1993. Luminosities fitted
over the age range $10^7\mbox{y}<t<10^{10.4}\mbox{y}$. IMF slope $x$ 
(equation \protect\ref{Ltotdef})
relevant only for stars with masses greater than $1\,M_\odot$.}
\tablenotetext{c}{Ref 2: Leitherer \&\ Heckman 1995. Luminosities fitted
over the age range $10^{6.5}\mbox{y}<t<10^{8.5}\mbox{y}$. IMF slope $x$
(equation \protect\ref{Ltotdef})
relevant only for stars with masses greater than $2\,M_\odot$.}
\end{deluxetable}

\clearpage
\figcaption[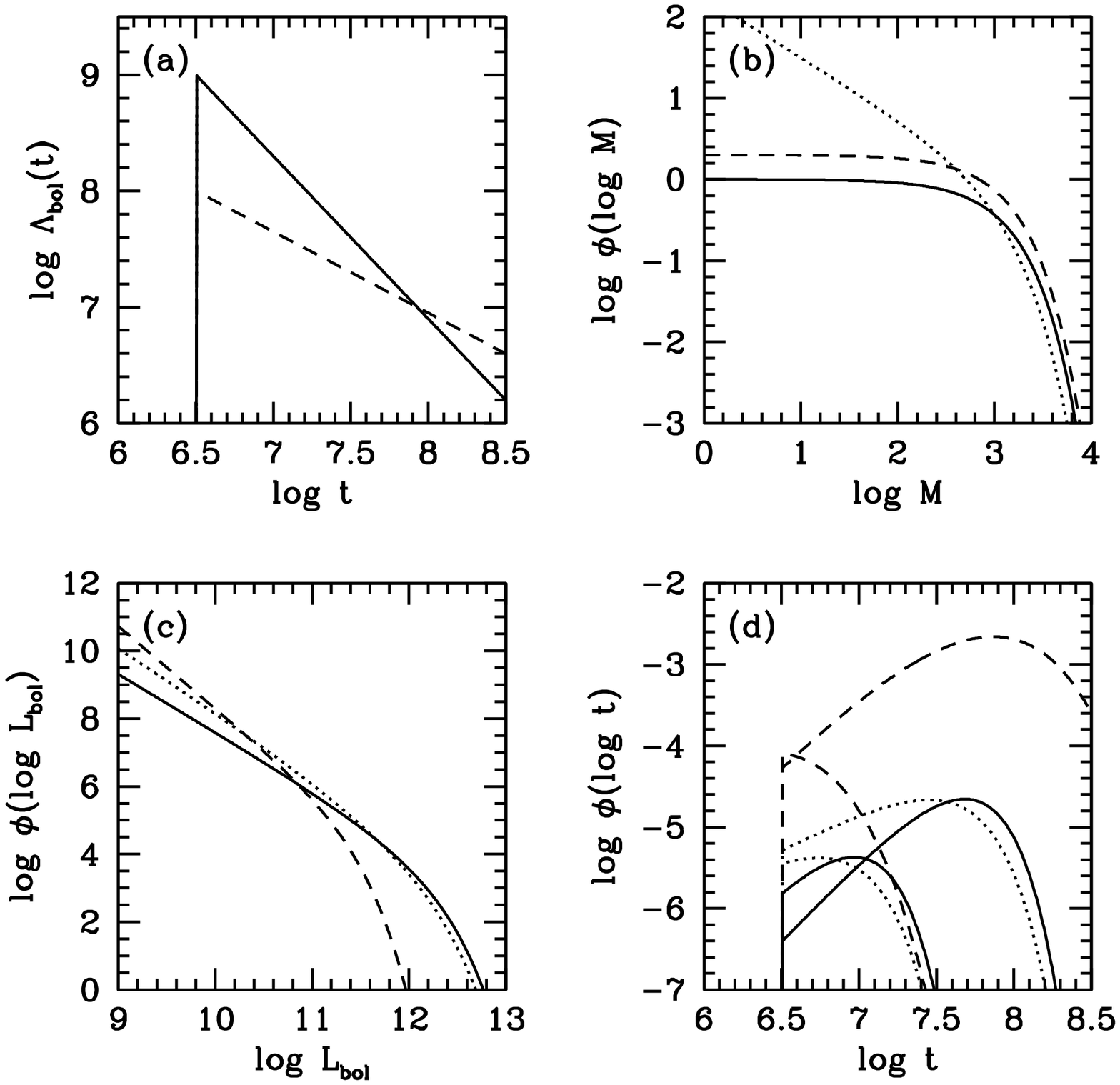]{
An illustration of the fading mechanism for the steep GLF presented in
this {\sl Letter.}  Shown are {\sl (a)}~the bolometric fading laws
$\Lambda_{\rm bol}(t)$ (bolometric luminosity per unit mass in solar
units) used, crude fits to Fig.~7 of Leitherer \& Heckman (1995); {\sl
(b)}~the mass functions $\phi(\log M)\propto M\,\phi(M)$ (number per
decade in mass), really the distributions of starburst masses,
arbitrarily choosing $M^{\ast}=10^6\,M_{\odot}$; {\sl (c)}~the
bolometric luminosity functions $\phi(\log L_{\rm bol})\propto L_{\rm
bol}\,\phi(L_{\rm bol})$ (number per decade in luminosity) produced by
the fading and mass functions; and {\sl (d)}~the age function
$\phi(\log t)\propto t\,\phi(t)$ (number per decade in age) for
objects at luminosity $L_{\rm bol}=10^{11.5}\,L_{\odot}$ (lower
curves) and $L_{\rm bol}=10^{10.5}\,L_{\odot}$ (upper curves).  Three
models are shown, one with $\zeta=-1.4$ and $\beta=-1.0$ (solid line),
$\zeta=-1.4$ and $\beta=-1.75$ (dotted), and $\zeta=-0.7$ and
$\beta=-1.0$ (dashed).  See text for symbol definitions.  Mass,
luminosity and age functions all have arbitrary normalizations; it is
not meaningful to compare relative heights of two different curves on
the same plot.
\label{fig1}}

\plotone{steeplf.eps}

\end{document}